\documentclass[amsmath,amssymb,superscriptaddress,prd,nofootinbib,twocolumn]{revtex4-2}
\usepackage{graphicx}
\usepackage{amsmath}
\usepackage{amssymb}
\usepackage{amsfonts}
\usepackage{enumerate}
\usepackage{color}
\usepackage{dsfont}
\usepackage{tikz}
\usepackage{ragged2e}
\usetikzlibrary{calc,decorations.markings}
\usepackage[colorlinks=true,linkcolor=blue,citecolor=blue,urlcolor=blue]{hyperref}
\usepackage{tipa}
\usepackage{soul}
\usepackage{float}
\usepackage{subfig}
\newcommand{\kk}{\boldsymbol{k}}
\newcommand{\xx}{\mathbf{x}}
\newcommand{\x}{\mathsf{x}}

\usetikzlibrary{decorations.pathmorphing, patterns,shapes}
\usepackage{comment}
\begin{document}

\title{Entangled states dynamics of moving two-level atoms in a thermal field bath}
\date{\today}

\author{Nikolaos Papadatos}\email{n.papadatos@upatras.gr}
\affiliation{Department of Physics, University of Patras, 26504 Patras, Greece}
\author{Dimitris Moustos}\email{dmoustos@upatras.gr}
\affiliation{Department of Physics, University of Patras, 26504 Patras, Greece}

%========================================================================================================================
\begin{abstract}
We consider a two-level atom that follows a wordline of constant velocity, while interacting with a massless scalar field in a thermal state through: (i) an Unruh-DeWitt coupling, and (ii) a coupling that involves the time derivative of the field. We treat the  atom as an open quantum system, with the field playing the role of the environment, and employ a master equation to describe its time evolution. We study the dynamics of entanglement between the moving atom and a (auxiliary) qubit at rest and isolated from the thermal field. We find that in the case of the standard Unruh-DeWitt coupling and for high temperatures of the environment the decay of entanglement is delayed due to the atom's motion. Instead, in the derivative coupling case, the atom's motion always causes the rapid death of entanglement.

\end{abstract}

\maketitle

%===========================================================================================================
\section{Introduction}

Quantum entanglement \cite{entangl:review}, which describes the non-local correlations between different quantum systems, is considered one of the most striking features of quantum theory.  The past few years, entanglement has been recognised as a valuable resource for quantum information processing. It lies at the heart  of many applications, such as teleportation \cite{Bennett:Teleport, teleportation},  cryptography \cite{Ekert,q:crypt}, and communication through dense coding \cite{dense:coding},  while it is necessary for the exponential speed–up of quantum computations over classical ones \cite{nielsen,Quantum:algo}.

However, in realistic conditions, the inevitable 
interaction of quantum systems with their surrounding environments results in a rapid loss, and an eventual death of entanglement at finite times (see, e.g., \cite{Open:system:entangl} and references therein). This fragility of entanglement has become one of
the main issues in the realization of practical quantum technologies. To protect entangled states from the  decoherence induced by the system-environment interactions, various techniques  have  been developed in the past, including dynamical decoupling \cite{Dynamical:decoupl}, the use of the quantum Zeno effect \cite{Zeno, Zeno:decoh} and decoherence-free subspaces \cite{Lidar:subspace}. 

Relativistic quantum information (RQI) combines the mathematical framework of quantum field theory in both flat and curved spacetime with the methodologies and concepts of quantum information theory and quantum optics. Its objective is to investigate the implications of relativistic motion, gravity and the curvature of spacetime on quantum information processes and quantum resources. To fulfill this objective, an important tool frequently employed in RQI is the Unruh-DeWitt (UDW) particle detector model \cite{Unruh,DeWitt,Hu:louko}. A UDW detector consists of a quantum system (such as a two-level atom or a harmonic oscillator) that locally interacts with a relativistic quantum field in a background spacetime, while being allowed to move along any spacetime trajectory. Initially introduced as an operationally way to probe the particle phenomenology in the celebrated Unruh and Hawking effects \cite{Unruh,birrell}, the UDW detector model has since found many applications in implementing quantum information protocols in relativistic settings, including relativistic quantum communication  \cite{Cliche,jonsson2017quantum,landulfo,jonsson:com:bh, tjoa:comm,Lapponi} and teleportation \cite{Alsing,PaulAlsing,hotta,landulfo:matsas}. 

Recent research in RQI has focused on exploring entanglement harvesting protocols \cite{VALENTINI1991,reznik,Reznik2,Perche}, which aim to extract the entanglement inherently present in a relativistic quantum field, onto initially uncorrelated UDW detectors through their interaction with it. The relativistic motion of detectors has been demonstrated to significantly influence the process of entanglement harvesting (see, e.g., \cite{Harvesting,Harv:circ,Liu:Mann,Barman,Bozanic}), where, for instance, relative inertial motion between two detectors, can, in some cases, increase the amount of extracted entanglement \cite{Harv:inert}. Similarly, the amount of coherence generated into a detector moving at constant speed while interacting with a  quantum field in a coherent state has been shown to exceed that of a static detector case \cite{NK,NK:catal}.  We then pose the following question: \emph{Can we harness relativistic motion to mitigate entanglement loss in dissipative systems?}

To address this issue, in the present paper, we investigate the implications of an atom's motion on the rate at which an initial amount of entanglement is lost as a result of the atom's interaction with a thermal environment. In particular, we employ the methods introduced in \cite{Doukas}, and study the dynamics of entanglement between two two-level atoms, with the one moving at a constant speed through a massless scalar quantum field in a thermal state, while the other is at rest and isolated from the field. We use two different coupling models to describe the atom-field interaction: (i) the standard UDW particle detector model, which linearly couples the atom with the field, and (ii) a derivative coupling model that couples the atom with the time derivative of the field \cite{Hinton,Grove,Takagi,BJA,DM}.  We find that in the first case, the motion of the atom can slow the rate at which entanglement is lost due to its interaction with a high temperature environment. This suggests the  use of an atom's (relativistic) motion as a method to counter entanglement loss in quantum systems. On the other hand, in the derivative coupling case, we observe that an increase in the atom’s velocity causes a rapid degradation of entanglement.

%=====================================================================================================================================

\section{Two-level atom moving through a thermal field bath}

We begin our analysis by introducing the UDW detector model. We then treat a detector that is moving with constant velocity as an open quantum system \cite{breuer}, with a scalar quantum field in a thermal state playing the role of the environment. We  employ a Markovian master equation to describe its time evolution.

Throughout the paper we denote spatial vectors with boldface letters $(\xx)$, while spacetime vectors are represented by sans-serif characters $(\x)$. We use the signature $(-+++)$ for the Minkowski spacetime metric. Unless otherwise specified we hereafter set $\hbar=c=k_B=1$.

%================================================================================================================
\subsection{The Unruh-DeWitt detector model}

We consider a  massless scalar quantum  field $\hat{\phi}(\x)$ in the (3+1)-dimensional Minkowski spacetime with metric $\eta_{\mu\nu}$. The scalar field satisfies the Klein-Gordon equation $\square \hat{\phi}(\x)=0$, where $\square=\eta^{\mu\nu}\partial_{\mu}\partial_{\nu}=-\partial_t^2+\nabla^2$ is the d'Alembert operator. In terms of plane wave  solutions to the Klein-Gordon equation the field can be expressed as
\begin{equation}\label{field}
    \hat{\phi}(t,\xx)=\int\frac{d^3\kk}{\sqrt{(2\pi)^3 2|\kk|}}\left(\hat{a}_{\kk}e^{-i(|\kk|t-i\kk\cdot\xx)}+\text{H.c.}\right),
\end{equation}
where $\hat{a}_{\kk}$ and $\hat{a}_{\kk}^\dagger$ are the annihilation and creation operators of the field mode with momentum $\kk$. They satisfy the  canonical commutation relations
\begin{align}\label{commut}
   [\hat{a}_{\kk},\hat{a}_{\kk'}^\dag]=\delta^{3}(\kk-\kk'), \quad [\hat{a}_{\kk},\hat{a}_{\kk'}]=[\hat{a}_{\kk}^\dag,\hat{a}_{\kk'}^\dag]=0
    .
\end{align}
The free Hamiltonian of the field, after subtracting the infinite zero-point energy, reads 
\begin{equation}
    \hat{H}_{\phi}=\int d^3\kk|\kk|\hat{a}_{\kk}^\dag\hat{a}_{\kk}.
\end{equation}

We consider an Unruh-DeWitt (UDW) particle detector \cite{Unruh,DeWitt,birrell,Hu:louko} modeled as a pointlike two-level atom (qubit) with energy gap $\omega$ and Hamiltonian
\begin{equation}\label{UDW:Hamiltonian}
    \hat{H}_0=\frac{\omega}{2}\hat\sigma_3,
\end{equation}
where $\hat{\sigma}_3$ is the usual Pauli operator. The detector moves along a worldline $\x(\tau)=(t(\tau),\xx(\tau))$ parametrized by its proper time $\tau$, while interacting with the scalar field through the interaction Hamiltonian
\begin{equation}\label{UDW:Hamiltonian}
    \hat{H}_{\text{int}}(\tau)=\lambda(\tau) \hat\mu(\tau)\otimes \frac{d^p}{d\tau^p}\hat{\phi}(\x(\tau)),
\end{equation}
where $\lambda(\tau)$ describes how the coupling between the detector and
the field is switched on and off, 
\begin{align}
	\hat{\mu}(\tau)=e^{i\omega\tau}\hat{\sigma}_++e^{-i\omega\tau}\hat{\sigma}_-
\end{align}
is the detector's monopole moment operator expressed in terms of the SU(2) ladder operators $\hat\sigma_{\pm}$, $\hat{\phi}(\tau):=\hat{\phi}(\x(\tau))$ is the scalar field operator evaluated along the detector's trajectory, and $p$ is a non-negative integer. We will next consider a sudden switching
$\lambda(\tau) = \lambda\theta(\tau)$, where $\lambda$ specifies the detector-field coupling strength and $\theta(\tau)$ is the Heaviside step function.

In this work, we will focus on the cases: (i) $p=0$, which  refers to the standard  UDW detector model, and (ii) $p=1$, which corresponds to a derivative coupling detector model \cite{Hinton,Grove,Takagi,BJA,DM} that linearly couples the atom with the proper-time derivative of the quantum field.  The derivative coupling model closely resembles the dipole interaction $\hat{\mathbf{d}}\cdot\hat{\mathbf{E}}$ between an atom with dipole moment $\hat{\mathbf{d}}$ and an external electromagnetic field $\hat{\mathbf{E}}(t,\boldsymbol{x})=-\partial_t \hat{\mathbf{A}}(t,\boldsymbol{x})$ expressed in terms of the vector potential $\hat{\mathbf{A}}$ (in the Coulomb gauge) \cite{Scully}.

In the followings, we will consider that the detector is moving with a constant velocity $\upsilon$ along the x-axis, i.e, following the trajectory \cite{birrell}
\begin{equation}\label{worldline}
    \x(\tau)=(\gamma\tau,\gamma\upsilon\tau,0,0),
\end{equation}
where $\gamma=1/\sqrt{1-\upsilon^2}$ is the Lorentz factor.

%===========================================================================================================

\subsection{The Markovian master equation}

The interaction Hamiltonian \eqref{UDW:Hamiltonian} constitutes a special case of the spin-boson model Hamiltonian \cite{breuer}. This implies considering the UDW detector as an open quantum system with the scalar field playing the role of the environment that induces dissipation and decoherence. We suppose that initially the field is in a thermal state 
\begin{equation}
    \hat\rho_{\beta}=\frac{e^{-\beta\hat{H}_{\phi}}}{\text{tr}(e^{-\beta\hat{H}_{\phi}})}
\end{equation}
at temperature $T=\beta^{-1}$, and that no correlations between the detector and the field  environment are present, i.e.,  $\hat{\rho}_{\text{tot}}(0)=\hat\rho(0)\otimes\hat\rho_{\beta}$.

The dynamics of the density operator of the combined detector-field system in the interaction picture is described by the Liouville-von Neumann equation
\begin{equation}\label{von:Neum}
    \frac{d}{d\tau}\hat{\rho}_{\text{tot}}(\tau)=-i[\hat{H}_{\text{int}}(\tau),\hat{\rho}_{\text{tot}}(\tau)],
\end{equation}
where $[\cdot,\cdot]$ stands for the commutator. We can formally solve Eq. \eqref{von:Neum} by integration to obtain
\begin{align}\label{von:Neum:int}
    \hat{\rho}_{\text{tot}}(\tau)&=\hat{\rho}_{\text{tot}}(0)-i\int_0^{\tau}ds[\hat{H}_{\text{int}}(s),\hat{\rho}_{\text{tot}}(s)].
\end{align}
Inserting then the integral form \eqref{von:Neum:int} back into \eqref{von:Neum} and taking the partial trace over the field bath degrees of freedom  yields an integro-differential equation for the reduced density matrix of the detector
\begin{equation}\label{reduced:eom}
     \dot{\hat{\rho}}(\tau)=-\int_0^{\tau}ds\,\text{tr}_{\phi}[\hat{H}_{\text{int}}(\tau),[\hat{H}_{\text{int}}(s),\hat\rho_{\text{tot}}(s)]].
\end{equation} 

We next successively  employ: (i) the \emph{Born approximation}, that is we assume a weak coupling between the detector and the field so that the combined state $\hat{\rho}_{\text{tot}}(s)\simeq\hat\rho(s)\otimes\hat\rho_{\beta}$ remains approximately factorized at all times, and (ii) the \emph{Markov approximation}, by which we replace $\hat\rho(s)$ by $\hat\rho(\tau)$ in the integral in \eqref{reduced:eom} and extend the upper limit of integration to infinity. The Markov approximation is valid provided that the detector's time evolution is much slower than that of the field bath \cite{nonMarkov:review}. After performing a change of variables $s\to\tau-s$ we obtain the Markovian master equation 
\begin{equation}
    \dot{\hat{\rho}}(\tau)=-\int_0^{\infty}ds\,\text{tr}_{\phi}[\hat{H}_{\text{int}}(\tau),[\hat{H}_{\text{int}}(\tau-s),\hat\rho(\tau)\otimes\hat\rho_{\beta}]].
\end{equation}

In terms of the interaction Hamiltonian \eqref{UDW:Hamiltonian} the above master equation takes the form \cite{DM:nonMarkov,DMmarkov}
\begin{align}\label{master}
\dot{\hat{\rho}}(\tau)=-\lambda^2\int_0^{\infty}ds\Big([\hat{\mu}(\tau),\hat{\mu}(\tau-s)\hat\rho(\tau)]\mathcal{W}(\tau,\tau-s)+\text{H.c.}\Big),
\end{align}
where 
\begin{equation}\label{wightman}
    \mathcal{W}(\tau,\tau'):=\text{tr}_{\phi}\Big(\hat\rho_{\beta}\hat\varphi(\tau)\hat\varphi(\tau')\Big)
\end{equation}
is the Wightman two-point correlation function of the field evaluated along the detector's trajectory; for convenience we have set $\hat\varphi(\tau)=\frac{d^p}{d\tau^p}\hat{\phi}(\tau)$. We note that when the state of the field is stationary and the detector follows a stationary spacetime trajectory \cite{Letaw}, such is the one with constant velocity, the Wightman function depends only on the proper time deference $\tau-\tau'$ between the two points on the detector's worldline and it can be expressed as $\mathcal{W}(\tau,\tau')=\mathcal{W}(\tau-\tau')$.

In the case of the standard UDW coupling, the Wightman function pull-backed to the inertial trajectory \eqref{worldline} followed by the detector takes the form (see Appendix \ref{appendix} for details)
\begin{align}\label{inert:wight}
\mathcal{W}_{\text{\tiny{UDW}}}(s)=&-\lim_{\epsilon\to 0^+}\frac{1}{4\pi^2(s-i\epsilon)^2}\nonumber\\ & \quad +\frac{\sqrt{1-\upsilon^2}}{4\pi^2\upsilon s}\int_{0}^{\infty} dk\,n_k\left[\sin\left(\sqrt{\frac{1+\upsilon}{1-\upsilon}}ks\right)\right.\nonumber\\&\quad\quad\quad\quad\quad\quad-\left.\sin\left(\sqrt{\frac{1-\upsilon}{1+\upsilon}}ks\right)\right]
\end{align}
where $n_k=(e^{\beta|\kk|}-1)^{-1}$ denotes the Planck distribution,  and the limit $\epsilon\to0^+$ is understood in the distributional sense. The Wightman function \eqref{inert:wight} is stationary. The first term is independent of the temperature and corresponds to the vacuum two-point correlation function of the field. The second term incorporates the thermal effects of the field environment.
On the other hand, in the case of the time-derivative coupling the Wightman function is
\begin{align}
    \mathcal{W}_{\text{\tiny{TD}}}(s)=-\frac{\partial^2}{\partial s^2}\mathcal{W}_{\text{\tiny{UDW}}}(s).
\end{align}

Markov approximation implies that the correlation function of the environment decays rapidly with respect to some characteristic timescale $\tau_c$. For an atom at rest, with worldline $\x(\tau)=(\tau,0,0,0)$, and interacting with the thermal field bath, this timescale is determined by $\tau_c=T^{-1}$. The correlation function is expressed as (see Appendix \ref{appendix})
\begin{align}
    \mathcal{\mathcal{W}}_{\text{\tiny{UDW}}}(s)=-\frac{1}{4\beta^2\sinh^2(\pi s/\beta) },
\end{align}
exhibiting a sharp peak around $s=0$ and  decaying exponentially as $s/\tau_c>>1$, i.e., $\mathcal{W}(s)\sim e^{-s/\tau_c}$. Thus, as commonly understood \cite{breuer,nonMarkov:review}, the Markov approximation fails to hold at low temperatures of the environment. Besides, as demonstrated in \cite{Papadatos2020}, an atom moving inertially at a constant speed $\upsilon$ and interacting with the thermal field bath, experiences a continuum of thermal  baths with temperatures $T'$ within the range
\begin{equation}
    T\frac{1-\upsilon}{1+\upsilon}\leq T'\leq T\frac{1+\upsilon}{1-\upsilon}.
\end{equation}
Consequently, in the ultra-relativistic limit $\upsilon\to1$, the moving atom can perceive the field bath as if it were in its vacuum (zero-temperature) state. This observation is also supported by the correlation function Eq.  \eqref{inert:wight} (or equivalently Eq. \eqref{thermal:two:point}).
For a fixed temperature, the correlation function $\mathcal{W}(s)$ remains sharply peaked around $s=0$ and exponentially decays with respect to the temperature. In the ultra-relativistic limit $\upsilon\to1$, the term blue-shifted by the Doppler factor $\sqrt{\frac{1+\upsilon}{1-\upsilon}}$ exponentially decays as $\sim \exp^{-T\sqrt{\frac{1+\upsilon}{1-\upsilon}}s}$, but instead the term red-shifted by the factor $\sqrt{\frac{1-\upsilon}{1+\upsilon}}$ behaves as $\sim \frac{1}{T\sqrt{\frac{1-\upsilon}{1+\upsilon}}s}$.  Therefore, we can conclude that in the ultra-relativistic regime the Markov approximation ceases to hold and a more thorough investigation that takes into account memory effects is required.

Calculating the Fourier transform time integrals in \eqref{master}, employing the (post-trace) rotating wave approximation (RWA) \cite{RWA,RWA:Agarwal} by which the rapidly oscillating terms of the form $\hat\sigma_+\hat{\rho}\hat\sigma_+e^{2i\omega\tau}$ and $\hat\sigma_-\hat{\rho}\hat\sigma_-e^{-2i\omega\tau}$ are ignored, and lastly transforming back to the Schrödinger picture, we can write (see Ref. \cite{Papadatos2020} for a detailed derivation)  the master equation in the \emph{Gorini-Kossakowski-Sudarshan-Lindblad} (GKSL) \cite{Lindblad,GKS}  form 
\begin{align}\label{Lindbland}
    \dot{\hat{\rho}}=&-i\frac{\Omega}{2}[\hat{\sigma}_3,\hat{\rho}]+\Gamma(N+1)\left(\hat\sigma_-\hat{\rho}\hat\sigma_+-\frac{1}{2}\{\hat\sigma_+\hat\sigma_-,\rho\}\right)\nonumber\\
    &\quad \quad \quad+\Gamma N\left(\hat\sigma_+\hat{\rho}\hat\sigma_--\frac{1}{2}\{\hat\sigma_-\hat\sigma_+,\rho\}\right),
\end{align}
where $\{\cdot,\cdot\}$ stands for the anticommutator, we have defined $\Omega\equiv\omega+\Delta\omega$ with $\Delta\omega$ being the frequency (Lamb) shift induced by the field environment, and  $\Gamma$ is the decay constant of the qubit detector in the vacuum. We note that the evolution map $\Phi:\rho(0)\mapsto\rho(\tau)$ that the GKSL master equation generates
is a completely positive and trace-preserving (CPTP) map \cite{breuer}. 

In the case of the UDW coupling the decay coefficient reads 
\begin{align}
    \Gamma_{\text{\tiny{UDW}}}=\frac{\lambda^2\omega}{2\pi},
\end{align}
whereas when considering the derivative coupling it takes the form
\begin{align}\label{Gamma:TD}
    \Gamma_{\text{\tiny{TD}}}=\frac{\lambda^2\omega^3}{6\pi}\left(1+2\cosh(2\text{arctanh}(\upsilon))\right).
\end{align}
Note that when the detector is static ($\upsilon=0$), the UDW coupling and derivative coupling cases produce equivalent expressions for the decay constant $\Gamma$ (with the exception of an additional factor of $\omega^2$ arising from the derivative nature of the latter coupling). Nonetheless, this is not the case when the detector is in constant motion ($\upsilon\neq0$). Accordingly, the coefficient $N$ is equal to
\begin{align}\label{N:udw}
	N_{\text{\tiny{UDW}}}=\frac{\sqrt{1-\upsilon^2}}{2\upsilon\beta\omega}\log\left(\frac{1-e^{-\beta\omega\sqrt{\frac{1+\upsilon}{1-\upsilon}}}}{1-e^{-\beta\omega\sqrt{\frac{1-\upsilon}{1+\upsilon}}}}\right)
\end{align}
in the case of the standard UDW coupling, and equal to 
\begin{align}
    N_{\text{\tiny{TD}}}  =& \frac{3(1-\upsilon^2)^{\frac{3}{2}}}{2\upsilon\beta^3 \omega^3  (3+\upsilon^2)}\times\nonumber\\ &\left[F\left(\beta \omega \sqrt{\frac{1-\upsilon}{1+\upsilon}}\right) - F\left(\beta \omega \sqrt{\frac{1+\upsilon}{1-\upsilon}}\right)\right],
\end{align}
in the case of the derivative coupling, where the function
\begin{equation}
    F(x) = 2\text{Li}_3(e^{-x}) + 2x \text{Li}_2(e^{-x}) + x^2\text{Li}_1(e^{-x})
\end{equation}
is expressed in terms of the polylogarithm function \cite{NIST}
\begin{equation}
    \text{Li}_s(z)=\sum\limits_{k=1}^{\infty}\frac{z^k}{k^s},
\end{equation}
which appears in the Bose-Einstein integral
\begin{equation}
    \text{Li}_{s+1}(e^x)=\frac{1}{\Gamma(s+1)}\int_0^\infty dk \frac{k^s}{e^{k-x}-1}.
\end{equation}
Note that in the static detector limit $\upsilon\to0$ the usual thermal Planck distribution
\begin{align}
    N=\frac{1}{e^{\beta\omega}-1}
\end{align}
that describes the mean number of field modes is recovered in both coupling cases. This implies that the atom's motion changes the way it experiences the Planckian spectrum of the background thermal environment, causing it to respond differently to the black body spectrum.

We do not report here the expressions for the frequency shift $\Delta\omega$, as its exact form is not involved in our later calculations and results (see, e.g.,  Eq. \eqref{conc:final}). The analytic expressions for the Lamb shift in both coupling cases can be found in \cite{Papadatos2020}. For a more detailed discussion regarding the different outcomes produced by the two couplings in the context of moving detectors in a thermal field bath we refer the reader to \cite{Papadatos2020}.
%===========================================================================================================
%=============FIGURE========================================================================
\begin{figure*}
\centering
    \subfloat[$\beta\omega=5$ (UDW)]{\includegraphics[scale=0.46]{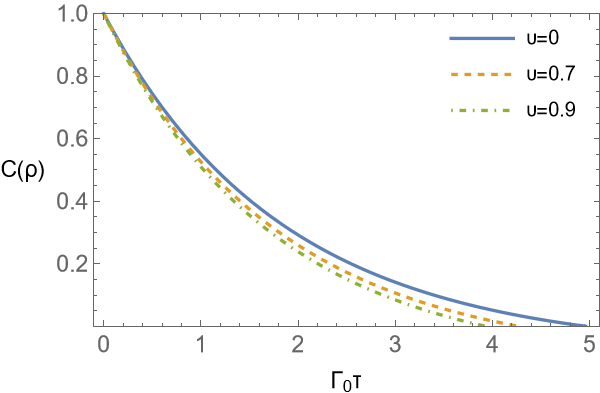}}  \hspace{0.1cm}
\subfloat[$\beta\omega=1$ (UDW)]{\includegraphics[scale=0.46]{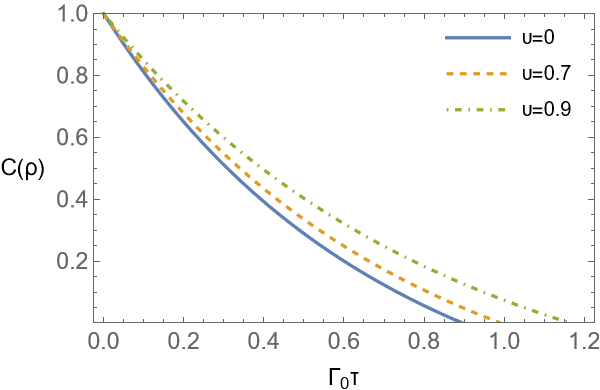}} \hspace{0.1cm}
\subfloat[$\beta\omega=0.5$ (UDW)]{\includegraphics[scale=0.46]{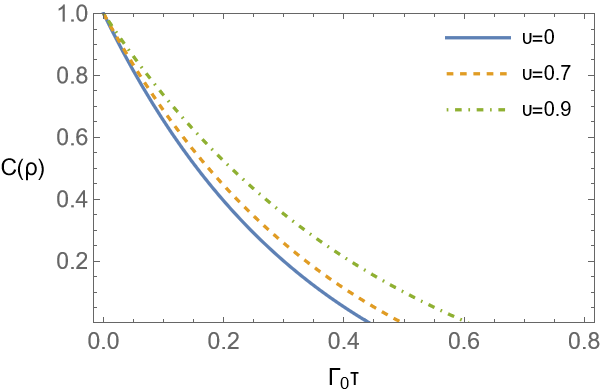}}\\
\subfloat[$\beta\omega=5$ (TD)]{\includegraphics[scale=0.46]{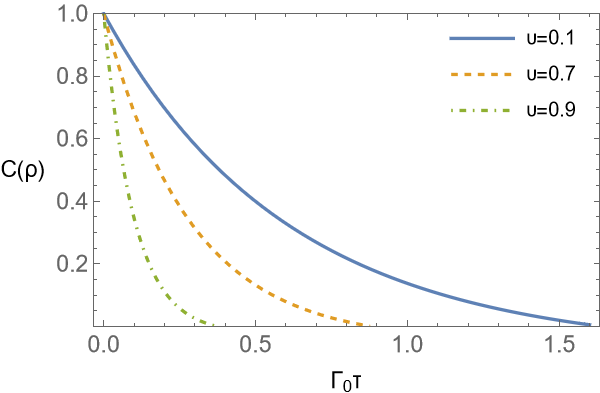}} \hspace{0.1cm}
\subfloat[$\beta\omega=1$ (TD)]{\includegraphics[scale=0.46]{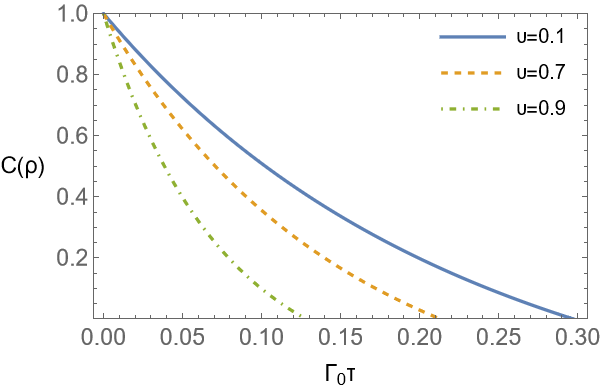}} \hspace{0.1cm}
\subfloat[$\beta\omega=0.5$ (TD)]{\includegraphics[scale=0.46]{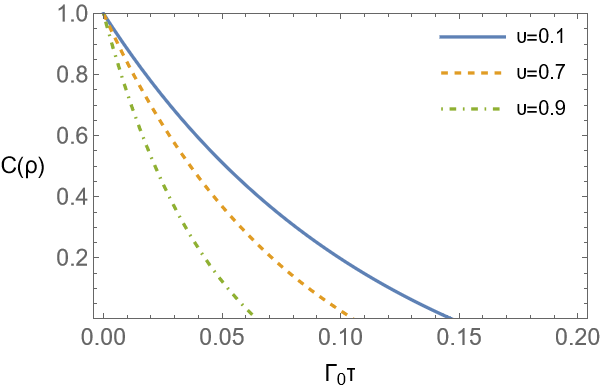}}
   \caption{ Concurrence as a function of time (in units $\Gamma_0^{-1}$) in the UDW (upper) and the derivative coupling case (lower) respectively, for different detector velocities $\upsilon$ and temperatures $T=\beta^{-1}$ (in units of $\omega$) of the field bath. For the UDW coupling $\Gamma_0=\frac{\lambda^2\omega}{2\pi}$, whereas for the derivative coupling $\Gamma_0=\frac{\lambda^2\omega^3}{6\pi}$.}
    \label{fig:conc}
\end{figure*}
%=============================================================================================

\section{Dynamics of entangled states}

 Motivated by \cite{Doukas,dS2014}, we next introduce an identical two-level detector as an auxiliary system that is initially entangled with the moving detector. The auxiliary system is kept isolated from the field environment and does not interact with it.

The density matrix of the two-atom system is expressed in terms of the Pauli matrices as
\begin{equation}\label{two-atom:state}
	\hat\rho(\tau)=\sum_{i=0}^{3}\sum_{j=0}^{3}u_{ij}(\tau)\hat\sigma_i \otimes \hat\sigma_j,
\end{equation}
where $u_{ij}=\frac{1}{4} \text{tr}(\rho \sigma_i \otimes \sigma_j)$ are the components of the Bloch vector, and $\hat{\sigma}_0=\hat1$ denotes the identity operator.
By substituting the density matrix \eqref{two-atom:state} into the master equation  \eqref{Lindbland} we obtain the following set of differential equations 
\begin{align}
    \dot{u}_{0j}(\tau)=&0,\nonumber\\
    \dot{u}_{1j}(\tau)=&-\frac{A}{2}u_{1j}(\tau)-\Omega u_{2j}(\tau),\nonumber\\
    \dot{u}_{2j}(\tau)=&-\frac{A}{2}u_{2j}(\tau)+\Omega u_{1j}(\tau),\nonumber\\
    \dot{u}_{3j}(\tau)=&-Au_{3j}(\tau)-Bu_{0j}(\tau)
\end{align}
for the coefficients $u_{ij}$, where for convenience we have defined 
    $A:=\Gamma(2N+1),$ and  $B:=-\Gamma$.
 Their solution reads
\begin{align}
    u_{0j}(\tau)&=u_{0j}(0),\nonumber\\
    u_{1j}(\tau)&=u_{1j}(0)e^{-\frac{A}{2}\tau}\cos(\Omega\tau)-u_{2j}(0)e^{-\frac{A}{2}\tau}\sin(\Omega\tau),\nonumber\\
    u_{2j}(\tau)&=u_{1j}(0)e^{-\frac{A}{2}\tau}\sin(\Omega\tau)+u_{2j}(0)e^{-\frac{A}{2}\tau}\cos(\Omega\tau),\nonumber\\
    u_{3j}(\tau)&=u_{3j}(0)e^{-A\tau}-u_{0j}(0)\frac{B}{A}\left(1-e^{-A\tau}\right).
\end{align}

We assume that initially the atoms are prepared in a maximally entangled Bell state
\begin{equation}
    \hat\rho(0)=\frac{1}{4}(\hat\sigma_0\otimes\hat\sigma_0+\hat\sigma_1\otimes\hat\sigma_1-\hat\sigma_2\otimes\hat\sigma_2+\hat\sigma_3\otimes\hat\sigma_3),
\end{equation}
that is the initial conditions $u_{00}(0)=u_{11}(0)=-u_{22}(0)=u_{33}(0)=1/4$ are applied. Thus, the evolution of the shared state between the moving atom and the auxiliary system is 
\begin{align}\label{shared:state}
    \hat\rho(\tau)=\frac{1}{4}\Bigg\{&\hat\sigma_0\otimes\hat\sigma_0+ e^{-\frac{A}{2}\tau}\cos(\Omega\tau)(\hat\sigma_1\otimes\hat\sigma_1-\hat\sigma_2\otimes\hat\sigma_2)\nonumber\\&+ e^{-\frac{A}{2}\tau}\sin(\Omega\tau)(\hat\sigma_2\otimes\hat\sigma_1+\hat\sigma_1\otimes\hat\sigma_2)\nonumber\\&+\frac{B}{A}(1-e^{-A\tau})\hat\sigma_3\otimes\hat\sigma_0+e^{-A\tau}\hat\sigma_3\otimes\hat\sigma_3\Bigg\}.
\end{align}

Our aim is to study how the amount of entanglement in the shared state \eqref{shared:state} is affected by the motion of the one detector partner. To this end, we next calculate its concurrence.

\subsection{Concurrence}

To quantify the amount of entanglement present in an arbitrary two-qubit state $\rho$ we employ the concurrence \cite{concur1,wootters:conc}\begin{equation}
    C(\rho)\equiv\text{max}\left(0,\lambda_1-\lambda_2-\lambda_3-\lambda_4\right), 
\end{equation}
where $\lambda_i$ are, in decreasing order, the square roots  of eigenvalues of the matrix $R=\rho(\sigma_y\otimes\sigma_y)\rho^*(\sigma_y\otimes\sigma_y)$. Here, $\rho^*$ denotes the complex conjugate of $\rho$ and $\sigma_y$ is the usual Pauli matrix. Concurrence ranges from $C=0$ for a state with no entanglement to $C=1$ for a maximally entangled state.

We note that in the case of the so-called X-states \cite{x:states,x:states2}, which are characterized by the fact that their density matrix has non-zero elements along the main diagonal and anti-diagonal, i.e.,
\begin{equation}\label{x:state}
\rho_X=
\begin{pmatrix}
\rho_{11} & 0 & 0 & \rho_{14}\\
0 & \rho_{22} & \rho_{23} & 0\\
0 & \rho_{23}^* & \rho_{33} & 0\\
\rho_{14}^* & 0 & 0 & \rho_{44}
\end{pmatrix},
\end{equation}
the concurrence is shown to be given by
\begin{equation}\label{x:state:conc}
   C(\rho_X)=2\,\text{max}\left(0,|\rho_{23}|-\sqrt{\rho_{11}\rho_{44}},|\rho_{14}|-\sqrt{\rho_{22}\rho_{33}}\right).
\end{equation}

 It can be easily shown that the state \eqref{shared:state} is a X-state of the form \eqref{x:state}. As a result, we can directly calculate concurrence by employing Eq. \eqref{x:state:conc}. We find that 
\begin{align}\label{conc:final}
	C(\rho)=\max{\big(0,e^{-\frac{A}{2}\tau}-\frac{1-e^{-A\tau}}{2A}\sqrt{A^2-B^2}\big)}.
\end{align}
The concurrence \eqref{conc:final} is plotted
in Fig. \ref{fig:conc} as a function of time, in the case of the standard UDW (plots (a)-(c)) and derivative (plots (d)-(f)) detector-field coupling, for various velocities $\upsilon$ of the moving detector and temperatures $\beta^{-1}$ of the field bath. As may be expected, the  initial amount of entanglement present in the shared state between the moving atom and its auxiliary partner monotonically decreases with time. In all cases, the rate of decrease becomes greater with the increase of the field bath temperature.

\begin{figure}
\centering
  \includegraphics[scale=0.66]{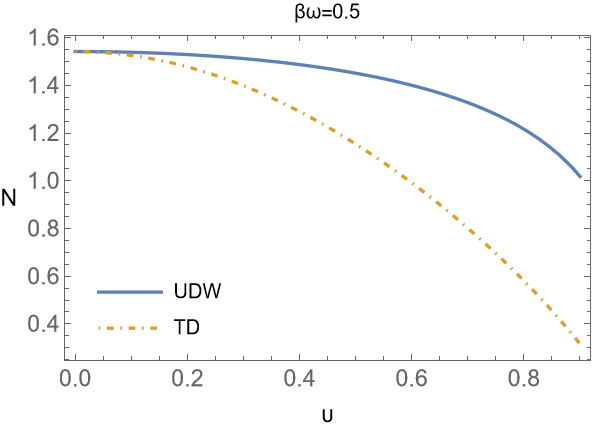}
   \caption{Coefficient $N$ as a function of the atom's speed $\upsilon$ for both UDW and derivative coupling cases, for a temperature $\beta\omega=0.5$ of the field bath.}
    \label{fig:Nhigh}
\end{figure}
\begin{figure}
\centering
\includegraphics[scale=0.66]{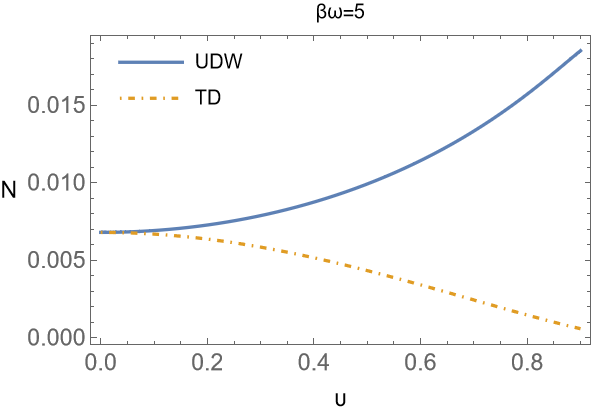}
   \caption{Coefficient $N$ as a function of the atom's speed $\upsilon$ for both UDW and derivative coupling cases, for a temperature $\beta\omega=5$ of the field bath.}  
    \label{fig:Nlow}
\end{figure}

Despite this fact, we observe that in the UDW coupling case and for temperatures $T\geq\omega$, entanglement loss is hindered when the atom is moving at a constant speed. This is a direct consequence of the way the moving atom responds to the perceived thermal spectrum $N$ of the environment. In Figs. \ref{fig:Nhigh} and \ref{fig:Nlow}, we plot $N_{\text{\tiny{UDW}}}(\beta,\upsilon)$ (Eq. \eqref{N:udw}) as a function of the atom's speed, for high and low temperatures of the field bath respectively. Concurrence approximately decreases exponentially as $C\sim e^{-\frac{\Gamma}{2}(2N+1)}$.  Besides, for higher temperatures of the environment, the coefficient $N$ behaves as
\begin{align}
    N_{\text{\tiny{UDW}}}=\frac{\sqrt{1-\upsilon^2}}{\upsilon\beta\omega}\log\sqrt{\frac{1+\upsilon}{1-\upsilon}}.
\end{align}
This means that $N_{\text{\tiny{UDW}}}$ decreases as a function of the atom's speed (see, also, Fig. \ref{fig:Nhigh}), consequently reducing the rate at which concurrence decreases.  In contrast, in the low-temperature regime, $N$ exponentially increases with the atom's speed (see, Fig. \ref{fig:Nlow}), behaving as
\begin{align}
    N_{\text{\tiny{UDW}}}=\frac{\sqrt{1-\upsilon^2}}{\upsilon\beta\omega}e^{\beta\omega\sqrt{\frac{1+\upsilon}{1-\upsilon}}}.
\end{align}
This leads to a greater loss of an initial amount of entanglement for a moving atom. In summary, a greater delay of the entanglement degradation can be achieved with an increase of the atom's speed. Although the higher the reservoir's temperature, the greater the loss of any initial amount of entanglement, the latter can be delayed by allowing the detector to move along a worldline of constant velocity. 

Instead, this conclusion does not hold true for the derivative coupling case, where  as the atom's speed increases, there is a substantial decrease in the concurrence. As it is evident from Figs. \ref{fig:Nhigh} and \ref{fig:Nlow}$,  N_{\text{\tiny{TD}}}$ monotonically decreases with the atom's speed in both the low and high-temperature regimes. However, unlike the UDW case, where the decay coefficient $\Gamma_{\text{\tiny{UDW}}}$ is independent of the detector's speed, in the derivative coupling case, $\Gamma_{\text{\tiny{TD}}}$ exponentially increases with the atom's speed (see, Eq. \eqref{Gamma:TD}). As a consequence,  entanglement rapidly decoheres as the atom's speed increases.

The different environments are classified according to the form of their spectral density $J(\omega)$, which usually follows the power-law  $J(\omega)\propto\omega^{\alpha}$\footnote{Physically, the spectral density should vanish for
very high values of frequencies. The pointlike detector case implies the introduction of a high frequency cut-off $\omega_c=\epsilon^{-1}$ as an exponential factor $e^{-\epsilon\omega}$ in the spectral density 
to regularize the high frequency behavior \cite{Schlicht_2004,Louko:Satz}.} \cite{breuer}. 
The most common choice for the exponent $\alpha$ is $\alpha=1$, in which case the
the environment is characterized as Ohmic. If  $\alpha<1$ the environment is characterized as subohmic, and if $\alpha>1$ as supraohmic respectively.   
We note that the massless scalar field bath  $\hat{\phi}(\x)$ represents an Ohmic environment (from Eq. \eqref{thermal:w} we have that $J(|\kk|)\propto|\kk|$), that is an environment whose spectral density is linear in frequency. Instead, when considering the derivative coupling case, the field bath $\dot{\hat\phi}(\x)$ acts as a supraohmic environment (from Eq. \eqref{therm:Wight:der} we have that $J(|\kk|)\propto|\kk|^3$). Thus, for higher values ($\beta\leq\omega^{-1}$) of temperature of an Ohmic environment, we conclude that an atom's motion can slow down the rate at which entanglement is lost, resulting in the preservation of entanglement for longer time. However, this is not true for supraohmic heat reservoirs: any interaction of moving atoms with them is always disastrous for entanglement.  

Let us note finally that the time point at which entanglement completely disappears (i.e., $C(\rho)=0$) is given by the expression
\begin{align}
    \tau=-\frac{2}{A}\ln\left[\frac{-A+\sqrt{2A^2-B^2}}{\sqrt{A^2-B^2}}\right].
\end{align}

%=========================================================================================
\section{Concluding remarks}

We demonstrated that by employing moving atoms one can decrease the rate at which entanglement is lost at high temperature Ohmic environments, consequently preserving entangled states for a longer period of time. Our result suggests the  use of an atom's (relativistic) motion as a potential method to mitigate entanglement loss in quantum systems, which is caused due to their unavoidable interaction with surrounding environments. We highlight that except for the fact that concurrence is frame dependent \cite{Robert}, it  also depends on the surrounding environment's characteristics. Besides, let us note that a similar to our result concerning  quantum coherence was reported in \cite{NK,NK:catal}, where the authors show that the rate at which coherence is lost, is sometimes slower for a moving atom than for an atom at rest. 

In future work, we aim to further explore the possibility of harnessing relativistic motion to counter entanglement loss in dissipative systems  by studying the dynamics of entanglement between two spatially separated atoms that are both moving through a common thermal environment.  Apart from reducing decoherence, it would be also interesting to investigate how the inertial motion of atoms affect the process of harvesting classical and quantum correlations from thermal quantum fields \cite{Harvesting:thermal}. We aim to tackle these questions in future work.
Finally, the relation between entanglement
generation and power of work extraction can be useful in quantum heat engines that take into  account relativistic effects \cite{Papadatos:Otto}.

%===========================================================================================
\appendix
\section{Thermal two-point function}\label{appendix}

We next evaluate the two-point correlation function $\mathcal{W}(\x,\x')=\text{tr}(\hat\rho_{\beta}\hat\phi(\x)\hat\phi(\x'))$ of a massless scalar field in a thermal state $\hat\rho_{\beta}$. Employing the identity 
\begin{align}
    \text{tr}\left(\hat\rho_{\beta}\hat{a}_{\kk}^\dagger\hat{a}_{\kk'}\right)=\frac{1}{e^{\beta|\kk|}-1}\delta^{(3)}(\kk-\kk')
\end{align}
and the canonical commutation relations \eqref{commut} we obtain
\begin{align}\label{thermal:w}
    &\text{tr}\Big(\hat\rho_{\beta}\hat\phi(\x)\hat\phi(\x')\Big)=\int\frac{d^3\kk}{(2\pi)^3 2|\kk|}e^{-i|\kk|(t-t')}e^{i\kk\cdot(\xx-\xx')}\nonumber\\&\quad+\int\frac{d^3\kk}{(2\pi)^3 2|\kk|}\frac{1}{e^{\beta|\kk|}-1}\left(e^{-i|\kk|(t-t')}e^{i\kk\cdot(\xx-\xx')}+\text{c.c.}\right).
\end{align}
Using spherical coordinates the Wightman function takes the form
\begin{align}\label{therm:Wight}
    &\mathcal{W}(\x,\x')=\mathcal{W}_{\text{vac}}(\x,\x')\nonumber\\
    &+\frac{1}{4\pi^2r}\int_0^{\infty}dk\,n_k\bigg(\sin\left(k(s+r)\right)-\sin\left(k(s-r)\right)\bigg),
\end{align}
where 
\begin{align}
    \mathcal{W}_{\text{vac}}(\x,\x')=-\lim_{\epsilon\to 0^+}\frac{1}{4\pi^2}\frac{1}{(s-i\epsilon)^2-r^2}
\end{align}
is the vacuum two-point correlation function, 
\begin{equation}
    n_k=\frac{1}{e^{\beta|\kk|}-1}
\end{equation}
is the Planck distribution and we have set $s=t-t'$ and $r=|\xx-\xx'|$. The integral in \eqref{therm:Wight} can be computed by using the formula \cite{gradshteyn2014table}
\begin{equation}
    \int_0^{\infty}\frac{e^{pk}-e^{qk}}{e^{\beta k}-1}dk=\frac{1}{\beta}\bigg(\psi\left(1-\frac{q}{\beta}\right)-\psi\left(1-\frac{p}{\beta}\right)\bigg),
\end{equation}
where $\psi(z)$ is the digamma (psi) function, along with the relation 
\begin{align}
 \Im\psi(1+ix)=-\frac{1}{2x}+\frac{\pi}{2}\coth(\pi x),
\end{align}
where $x\in \mathbb{R}$. Employing also the formula
\begin{align}
    \lim_{\epsilon\to 0^+}\frac{1}{x\pm i\epsilon}=\mp i\pi\delta(x)+\mathcal{P}\frac{1}{x},
\end{align}
where $\mathcal{P}$ denotes the Cauchy principal value, the thermal two-point function (see, also, \cite{weldon}) finally reads
\begin{align}\label{thermal:two:point}
    \mathcal{W}(\x,\x')&=\frac{i}{8\pi r}\bigg(\delta(s+r)-\delta(s-r)\bigg)\nonumber\\&+\frac{1}{8\pi r\beta}\bigg[\coth{\left(\pi\frac{s+r}{\beta}\right)}-\coth{\left(\pi\frac{s-r}{\beta}\right)}\bigg].
\end{align}
We note that in the limit $r\to0$ the thermal Wightman function reads
\begin{align}
    \mathcal{W}(\x,\x')&=-\frac{1}{4\beta^2\sinh^2(\pi s/\beta)}.
\end{align}

In the same way we obtain
\begin{align}\label{therm:Wight:der}
   &\text{tr}\Big(\hat\rho_{\beta}\dot{\hat\phi}(\x)\dot{\hat\phi}(\x)\Big)=-\int\frac{d^3\kk |\kk|}{2(2\pi)^3 }e^{-i|\kk|(t-t')}e^{i\kk\cdot(\xx-\xx')}\nonumber\\&\quad-\int\frac{d^3\kk|\kk|}{2(2\pi)^3}\frac{1}{e^{\beta|\kk|}-1}\left(e^{-i|\kk|(t-t')}e^{i\kk\cdot(\xx-\xx')}+\text{c.c.}\right),
\end{align}
which, bearing in mind Eq. \eqref{thermal:w}, can equivalently be written as
\begin{equation}
    \text{tr}\Big(\hat\rho_{\beta}\dot{\hat\phi}(\x)\dot{\hat\phi}(\x)\Big)=-\frac{\partial^2}{\partial s^2}\text{tr}\Big(\hat\rho_{\beta}\hat\phi(\x)\hat\phi(\x')\Big).
\end{equation}

%=====================================================================================
\bibliography{references}
\end{document}